\documentclass{epsconf}
\usepackage{graphicx}
\usepackage{epsfig} 
\usepackage{wrapfig}
\usepackage{amsmath}
\usepackage{amsfonts}
\usepackage{subcaption}
\title{Cross-field chaotic transport of electrons by E $\times$ B electron drift
instability in Hall thrusters}

\author{\underline{D. Mandal}$^{1,2}$,Y. Elskens $^1$ , N. Lemoine $^3$ and F. Doveil $^1$}
\institute{$^1$ Aix-Marseille Universit{\'e}, CNRS, UMR 7345-PIIM Laboratory, Marseille, France\\
$^2$ Indo-French Centre for the Promotion of Advanced Research-CEFIPRA, New Delhi, India\\
$^3$ Universit\'{e} de Lorraine, Institut Jean Lamour, UMR-7198, CNRS, France}
\begin{document}
\maketitle

\begin{abstract}
One special interest for
the industrial development of Hall thruster is characterizing the anomalous 
cross-field electron transport observed after the channel exit. Since the 
ionization efficiency is more than 90$\%$, the neutral atom density in that 
domain is so low that the electron collisions cannot explain the high
electron flux observed experimentally. Indeed this is 100 times higher than 
the collisional transport. 
In Hall thruster geometry, as ions are not magnetized the electric and 
magnetic field configuration creates a huge difference in drift velocity 
between electrons and ions, which generates electron cyclotron drift
instability or $\vec E \times \vec B$ electron drift instability. 
Here we are focusing on collision-less chaotic transport
of electrons by those unstable modes generated by $\vec E \times \vec B$ 
drift instability. We found that in presence of these electrostatic 
modes electron 
dynamics become chaotic. They gain energy from the background waves which 
increases electron temperature along perpendicular direction by a significant 
amount, $T_\perp/T_\parallel\sim 4$, and a significant amount of crossfield 
electron transport is observed along the axial direction.  
\end{abstract}

\subsection{Introduction and numerical model}
In Hall thruster geometry, the electric and magnetic field configuration creates a 
huge difference in drift velocity between electrons and ions, which generates 
electron cyclotron drift instability or $\vec E \times \vec B$ electron drift instability \cite{Tsikata}. Unstable modes generated from this instability
have an important role in cross-field anomalous transport of electrons. One special interest for the industrial development of Hall thruster is characterizing the 
anomalous cross-field electron transport observed after the channel exit. Since the ionization efficiency is more than 90$\%$, the neutral atom density in that domain is so low that the electron collisions cannot explain the high electron flux observed experimentally. Here we focus on collision-less chaotic transport
of electrons by the unstable modes generated by the $\vec E \times \vec B$  
drift instability.
These unstable modes can evolve at a sufficient level of turbulence into a 
non-magnetic ion-acoustic instability with modified angular frequency
given  \cite{Cavalier} by,
\begin{eqnarray}
1+ k^2\lambda_{\rm De}^2
+g\left(\frac{\omega-k_{y}v_{\rm d}}{\omega_{\rm ce}},(k_{x}^2+k_{z}^2)
\rho_{\rm e}^2,
k_{x}^2\rho_{\rm e}^2\right) -\frac{k^2\lambda_{\rm De}\omega_{\rm pi}}
{(\omega -k_{z}v_{\rm i,b})^2} = 0, 
\label{dispersion_rel}
\end{eqnarray}
where $\lambda_{\rm De}$ is the electron Debye length, $v_{\rm d}= E_z/B$ is 
the electron 
drift velocity, $v_{\rm{i,b}}$ is the ion beam velocity, 
$\rho_{\rm e}=v_{\rm the}/\omega_{\rm ce}$ is the electron Larmor radius,
$v_{\rm the}$ is the electron thermal velocity. 
We consider a Cartesian coordinate system, $x$-direction as magnetic field direction, $y$-direction as $\vec{E}\times \vec{B}$ drift direction and $z-$direction as constant electric field direction, which are the radial, azimuthal and axial direction of the thruster chamber, respectively.
$\omega$, $\omega_{\rm ce}$ and $\omega_{\rm pi}$ are the mode, electron cyclotron and ion plasma frequencies, 
respectively, and $g$ is the Gordeev function \cite{Gordeev}.
This analytical model for the dispersion relation fits well with the experimental data. We consider a constant electric field ${\rm E_{0}}\hat{z}$
along the $z$-direction and a constant magnetic field $\vec B = {\rm B_{0}} \hat{x}$ along $x$-direction.

Experimentally, the observed propagation angle of the instability-generated wave deviates by
$\tan^{-1} ({ k_z/k_y }) \sim 10-15^0$  from the
azimuthal $y-$direction near the thruster exit plane. Further from the exit plane, the propagation becomes progressively more azimuthal \cite{Tsikata}. Hence, the wave vector along axial direction ${ k_z \sim 0.2k_y}$, and the electric field along the axial direction is dominated by the stronger constant field ${\rm E_0}\hat{z}$.
Therefore for simplicity, we consider that the unstable modes are confined in $x-y$ 
(ie., $r$-$\theta$) plane only. Then the time varying part of the potential
in $x-y$ plane is constructed as a sum of unstable modes.
The total electric field acting on the particle is
\begin{eqnarray}
\overrightarrow{E}(x,y,z,t) &= &\sum_n {\rm \phi}_{0n} \left[{ k_{nx}} 
\sin \alpha_{n}(x,y,t) \hat{x} +  
{k_{ny}}  \sin \alpha_{n}(x,y,t) 
\hat{y} \right] + {\rm E_{0}} \hat{z},
\label{e-field}
\end{eqnarray} 
with the phase $\alpha_n(x,y,t) =  k_{nx} x +  k_{ny} y-{\rm \omega}_n t + {\rm \zeta}_n$, where $n$ is a label for different modes with
wave vector $\vec{ k}_n$, angular frequency ${\rm \omega_n}$ and 
 phase ${\rm \zeta_n}$. ${\vec{k}_n}$, ${\omega_n}$ follow the dispersion relation eq.~(\ref{dispersion_rel}) and phases  $\zeta_n$ are random. Here the position $\vec x$, velocity $\vec v$, 
 time $t$ and the potential ${\rm \phi_0}$ are normalized with Debye length 
 $\lambda_{\rm De}$, thermal velocity $v_{\rm the}$, electron plasma frequency $\omega_{\rm pe}^{-1}$ and $m_e v_{\rm the}^2/q$, respectively.
 We choose the amplitude ${\rm \phi}_{0n}$ of all 
the modes equal to the saturation potential at the exit plane of the thruster
$\vert \delta \phi_{y,{\rm rms}}\vert =T_{\rm {e}}/(6\sqrt{2})= 0.056 v_{\rm the}^2$ \cite{Boeuf}. We consider 
three modes $(n= 1, 2,3)$ with (${ k_{nx}, k_{ny}, \omega_n}$) =
($0.03, 0.75, 1.23\times 10^{-3}$), ($0.03, 1.5, 1.7\times 10^{-3}$) and 
($0.03, 2.25, 1.87\times 10^{-3}$), respectively. In normalized units, 
$q{\rm B_0}/m_{\rm e}= 0.1 \omega_{\rm pe}$, $q{\rm E_{0}}/m_{\rm e}= 0.04 \omega_{\rm pe} v_{\rm the}$, and ${ v_{\rm d} = 0.4 v_{\rm the}}$.
The equations of motion of the particle are
\begin{eqnarray}
 \frac{{\rm d}\vec{x}}{{\rm d}t} = \vec{v},
~~~
\frac{{\rm d}\vec{v}}{{\rm d}t} = \vec{E} + \vec{v}\times \vec{B}.
\label{eq_motion}
\end{eqnarray} 
Because $\vec E$ depends on space, the infinitesimal generators for both 
equations do not commute, and one uses a time-splitting numerical integration 
scheme. The first equation is integrated in the form 
$\vec x (t + \Delta t) = {\cal T}_{v, \Delta t} (\vec x(t)) =  
\vec x (t) + \vec v \Delta t$.
For the second equation, we separate the
 electric integration
$\vec v (t + \Delta t) = {\cal T}_{E, \Delta t} (\vec v(t)) = \vec v (t) + (q/m) \vec E \Delta t$
from the magnetic integration, which solves only the gyro-motion. 
For the latter, we use the Boris method \cite{boris}, formally
$\vec v(t + \Delta t) = {\cal T}_{B,\Delta t} \vec v (t)$.
As a result, we use a second-order symmetric scheme
\begin{equation}\nonumber
 \left( \begin{array}{c} x(t+\Delta t) \\ y(t+\Delta t) \end{array} \right) = {\cal T}_{v, \Delta t/2}\circ {\cal T}_{E, \Delta t/2} \circ{\cal T}_{B, \Delta t} \circ
{\cal T}_{E, \Delta t/2} \circ{\cal T}_{v, \Delta t/2}\left( \begin{array}{c} x(t) \\ y(t) \end{array} \right).
\end{equation}
\subsection{Time evolution of particle trajectory and velocity}
\begin{figure}[h!]
 \includegraphics[width=15 cm]{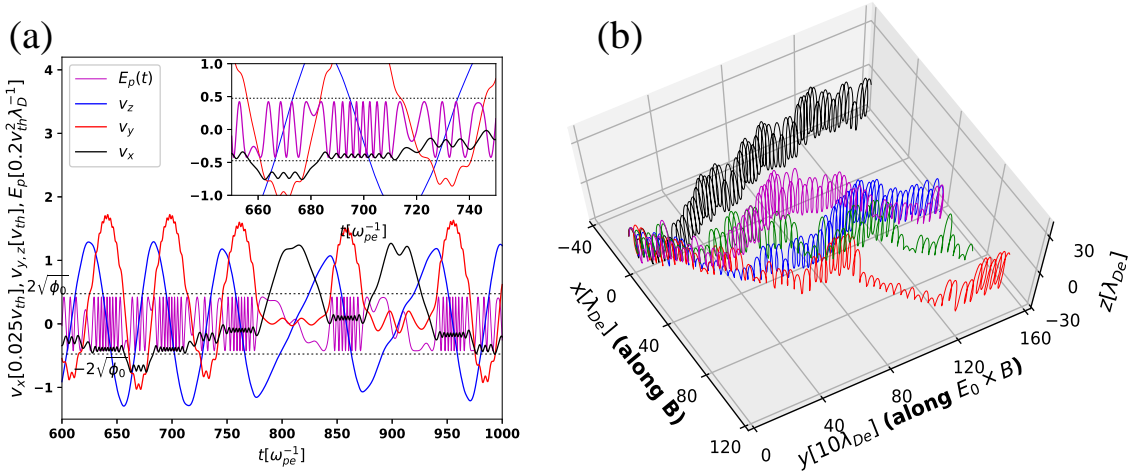}
 \caption{Particles evolution in the presence of a single background electrostatic wave with $n=2$.
Panel (a): velocity 
 components $v_x$ (black solid line), $v_y$ (red) and $v_z$ (blue) of one particle. Magenta line: electric field at particle location. Near $t=800$ and $900$, 
 the particle is trapped in the wave potential and it oscillates with the time 
 period $\tau_{\rm b} = 18 \omega_{\rm pe}^{-1}$. Panel (b): trajectories of 5 different particles with different 
 initial phase. }
 \label{Traj_vel}
\end{figure}

We solve the equation of motion Eq.~(\ref{eq_motion}) numerically for 1056 particles. 
In the absence of the background electrostatic waves 
$E_x = E_y =0$, their trajectories are regular and exhibit cyclotron motion with a drift
velocity ${ v_{\rm d}=0.4}$. Therefore, their velocity components are $v_{x}={\rm v}_{0x},
v_{y}= {\rm v_{\perp 0}} \cos({\rm \omega_c} t) + { v_{\rm d}}$ and 
$v_{z} = {\rm v_{\perp 0}} \sin({\rm \omega_c} t)$, where 
${\rm v_{\perp 0}} = \sqrt{{\rm v}_{0z}^2 +({\rm v}_{0y}- v_{\rm d})^2}$ and 
(${\rm v}_{0x}, {\rm v}_{0y}, {\rm v}_{0z}$) are the initial velocity components. In the presence 
of the background electrostatic wave, the wave-particle interaction modifies their cyclotron motion. The strength of the wave-particle interaction depends on the 
 wave amplitude and the particle velocity. Fig.~\ref{Traj_vel}(a) presents the time evolution 
of the three velocity components and the electric field $E_p(t)$ (magenta line)
at the particle location. Due to the cyclotron motion, $v_y$ oscillates about the drift 
velocity ${ v_{\rm d}}$ (solid red line). During each cyclotron oscillations,
when $\mid v_y \mid \leq 2\sqrt{\rm \phi_0}$ (denoted by black dashed lines)
the particle strongly interacts with the electrostatic wave, and the electric field 
$E_p(t)$ increases/decreases the $v_x$ value by large amount. The inset of Fig.~\ref{Traj_vel}(a) presents,
during strong interaction, according to the sign of $E_p$, jumps of $v_x$ (black solid line) in positive and negative direction.
Moreover, during this strong interaction depending on the local potential profile, the particle
may be trapped in the wave potential well and oscillate with the bounce-frequency
${\rm \omega_b = 0.35 \omega_{pe}}$. In Fig.~\ref{Traj_vel}(a) near $t=800$ and $900$, it is
trapped. One essential condition for the trapping is 
${\rm \omega_b> \omega_c }$, where $ \omega_{\rm b} = k_y {\rm \sqrt{q\phi_0/m}}$ is the
bounce frequency. Since $k_y \gg k_x$, the condition for trapping is easily satisfied along 
the $y-$direction,
therefore the particle bounces back and forth along $y-$direction and moves freely
along the $x-$direction. Hence along $x-$direction it 
gains/loses energy from/to the wave which causes a large change in $v_x$. Finally, depending
of the local potential value, it may escape from the wave and again start to 
exhibit cyclotron motion. Therefore the duration of trapping depends on $v_x$ 
and ${\rm \omega_b/ \omega_c }$. It is observed that, for small $v_x \ll \sqrt{\rm \phi_0}$,
this trapping is easily observed for ${\rm \omega_b/ \omega_c \geq 2}$.
 Outside the strong interaction region, due to the 
large particle velocity, electric field at particle location $E_p$ changes rapidly, 
which generates the small-amplitude fast oscillation in $v_x$. $v_y$ is also modulated 
due to this fast change in $E_p(t)$. Since the electric field along $z$-direction
${\rm E_{0}}\hat{z}$ is constant, the amplitude of the fast oscillation in $v_z$ is
negligible. The motion along $z-$direction is coupled with the other two 
directions due to $\vec{v}\times \vec{\rm B}$ term of Lorentz force, therefore $v_z$ is also modified during the strong interactions. 
In fig.~\ref{Traj_vel}(a) at $t= 900$,
during trapping, the oscillation of
$v_z$  is observed with frequency $\omega_{\rm b}$, on top of cyclotron motion. 

Fig.~\ref{Traj_vel}(b) presents the 
trajectories of 5 particles with slightly different initial phases. In the 
absence of the 
electrostatic wave, they exhibit cyclotron motion with drifting guiding center,  and their 
trajectories remain confined in $y-z$ plane.  
Due to the strong 
interaction with the electrostatic wave in presence of magnetic field, each 
trajectory evolves differently and separates exponentially from 
each other, and the dynamics become chaotic. During each strong interaction, there is a
change in the trajectories along $x$, and during trapping their average $y$ location remain unchanged. The duration of strong interaction depends ${\rm \omega_b/ \omega_c}$, therefore for single wave chaos will occurs for amplitude ${\rm \phi_0}$  satisfy $ \phi_{0} > \omega_{\rm c}^2/ k_y^2$. For
thruster parameter values, all three waves satisfy this criterion. In the presence of two and three waves,
the dynamics become more chaotic and this threshold value is redused.  
\subsection{Energy gain by the particles and their axial transport}
%
\begin{figure}
 \includegraphics[width=\linewidth]{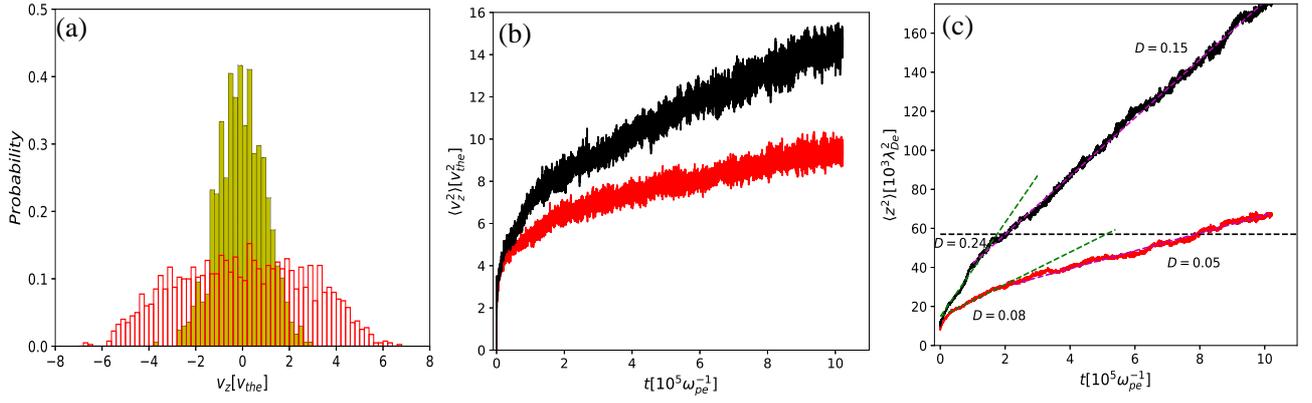}
 \caption{Panel(a): initial (yellow solid bar) and final (bar with red boundary) velocity distribution along $z$ at 
$t= 5\times 10^4\omega_c^{-1}$. 
Panels (b) and (c): mean square velocity dispersion  
$\langle v_z^2(t)\rangle$  and mean square displacement $\langle z^2(t)\rangle $,
 respectively. The red and black lines correspond to no-boundary and reflecting boundary cases, respectively
Panel (c) reveals two diffusion regimes in each curve, namely (0.08, 0.05) for no-boundary and (0.24, 0.15) for reflecting boundary. }
\label{Trans}
 \end{figure}
 To analyze transport, we consider 1056 particles with random initial positions in the rectangle $0\leq y_0 \leq 4\pi/ k_{1y}$, $0 \leq x_0 \leq 2\pi/ k_{1x}$ and with velocities drawn from a 3D Gaussian distribution with unit standard-deviation along all three directions. Then we evolve their dynamics in 
 presence of all three 
waves with equal amplitude $ \phi_{n0} = {\rm \phi_{0,rms}} $. For single wave interaction, the Hamiltonian 
of the dynamics can be written in a time independent form and therefore, though the dynamics remain chaotic, 
there is no net gain/loss of energy over long time evolution. But
in presence of two/three waves, the Hamiltonian is no more time independent, all the trajectories 
become chaotic and due to the wave particle interaction they gain energy from the waves. Their net 
perpendicular velocity $v_y, v_z$ increase. After sufficiently long time-evolution, they form a Gaussian-like velocity distribution profile
with higher temperature along $y-$ and $z-$directions. Since $E_x \ll E_{y,z}$,
the increase of the velocity component along the magnetic field is negligible compared to
the other two directions. Therefore the temperature along the magnetic field remains nearly unchanged. Fig.~\ref{Trans}(a) presents the initial ($t=0$) (solid yellow bars) and final $(t= 5\times 10^4\omega_c^{-1})$ (bars with red border) velocity distribution of $v_z$, which presents a significant increase of temperature along perpendicular direction $T_\perp $ compared to the parallel
direction, $T_{\perp}/T_{\parallel} \sim 4 $.

In the thruster chamber, there is an insulating boundary along $x-$direction. 
The width of the annular
space in the thruster is $240 \lambda_{\rm De}$. Therefore the particles are reflected when they reach 
to the boundary. If there were no reflection, particles would proceed under the same dynamics (red line in Fig.~\ref{Trans}(b)-(c)). To account for reflection (black line), we consider the Debye sheath electron potential energy near the 
wall to be ${\rm \phi_{sh} = 20 eV = 0.8 v_{the}^2}$. Electrons reaching 
the wall with $v_x < \sqrt{0.8}$ are specularly reflected, and electrons with 
$v_x > \sqrt{0.8}$ are isotropically reflected from the wall with conserving
their total energy. 
Fig.~\ref{Trans}(b)-(c) present 
$\langle v_z^2(t)\rangle $ and $\langle z^2(t)\rangle $ for reflecting boundary (black) and without boundary (red), where $\langle \rangle$ denotes the average over number of particles. The duration of strong interaction with the waves and hence 
the gain of energy from the waves decrease with increase of 
particle velocity. Therefore, the rate of energy gain in Fig.~\ref{Trans}(b)  
decreases with time for both cases. In isotropic reflection, the velocity components of the particle are redistributed randomly in three directions, a particle with small $v_y$ and $v_x$ gains more energy 
from the electrostatic wave compared to that having higher $v_y$ and $v_x$. Therefore, in presence of reflecting boundary, particles gain more energy than in absence of reflection. The dashed black line marks the location of thruster outlet along the $z-$direction. Since with reflection they gain more energy, their mean square displacement along $z-$direction crosses the thruster outlet, and they exit from the thruster chamber more quickly than in the case without boundary. For both cases, we found two different diffusion coefficient $(D= d\langle z ^2\rangle/dt)$, values, which are $(D = 0.08, 0.05)$ for no-reflection and 
$(D = 0.24, 0.15)$ for reflecting boundary. The change in slope around 
$t=2\times 10^5 \omega_{\rm pe}^{-1}$ is related to the structure of the stochastic web controlling 
the velocity transport \cite{zaslavsky,leoncini}.

Due to the chaotic dynamics, in presence of the single wave also we get a crossfield transport along $z$ direction, but the diffusion coefficient is very small. As the electric field $E_y$ is proportional to $k_y$, waves with different 
$\vec k$ values induce different diffusion coefficients and energy gain rates.  
\subsection{Conclusions}
Due to strong interaction with the wave potential, the drifted cyclotron motion becomes chaotic. In presence of more than one wave electrons gain energy over long time evolution and their temperature is increased. This chaotic dynamics helps in transport of electrons along the thruster axial direction. Significant amount of axial electron transport is observed in presence of more than one waves, and the electrons exit from the thruster chamber. The reflection at boundary enhances the transport coefficient.

\section*{Acknowledgements}
This work is part of IFCPRA project 5204-3. We acknowledge the financial support from CEFIPRA/IFCPRA. We are thankful to the Aix-Marseille University  MesoCentre for computations. We are grateful to Professors Xavier Leoncini, Dominique Escande and Abhijit Sen for many fruitful discussions and their comments. 

\end{document}